\documentclass[aps,prl,twocolumn,showpacs,superscriptaddress,preprintnumbers,amsmath,amssymb]{revtex4}

\usepackage{graphicx,color}
\usepackage{amssymb}
\usepackage{dcolumn}
\usepackage{bm}
\usepackage{hyperref}
\usepackage{color}
\usepackage{epstopdf}
\usepackage{times}
\begin{document} 

\title{Plasma Formation Dynamics in Intense Laser-Droplet Interaction}

\author{T.V.~Liseykina}\email{tatyana.liseykina@uni-rostock.de}\affiliation{Institut f\"ur Physik, Universit\"at Rostock, 18051 Rostock, Germany}
                       \affiliation{Institute of Computational Technologies SD RAS, 630090 Novosibirsk, Russia}
\author{D.~Bauer}\affiliation{Institut f\"ur Physik, Universit\"at Rostock, 18051 Rostock, Germany}

\begin{abstract}
We study the ionization dynamics in intense laser-droplet interaction using three-dimensional, relativistic particle-in-cell simulations. 
Of particular interest is the laser intensity and frequency regime for which initially transparent, wavelength-sized targets are not homogeneously ionized. Instead, the charge distribution changes both in space and in time on a sub-cycle scale. One may call this the extreme nonlinear Mie-optics regime.
We find that---despite the fact that the plasma created at the droplet surface is overdense---oscillating electric fields may 
penetrate  into the droplet under a certain angle, ionize, and propagate in the just generated plasma. This effect can be attributed to 
the local field enhancements at the droplet surface predicted by  standard Mie theory. The penetration of the 
fields into the droplet leads to the formation of a highly inhomogeneous charge density distribution in the droplet interior, 
concentrated mostly in the polarization plane. We present a self-similar, exponential fit of the fractional ionization degree 
which depends only on a dimensionless combination of electric field amplitude, droplet radius, and plasma frequency 
with only a weak dependence on the laser frequency in the overdense regime.

\end{abstract}


\maketitle

{\em Introduction ---}
Spherical, wavelength-sized, homogeneous dielectric or metal objects in plane-wave electromagnetic radiation fall into the realm of 
Mie theory \cite{Mie} and are of fundamental importance in optics.
Standard Mie theory provides the electromagnetic field configuration inside and outside a homogeneous sphere of a given dielectric 
constant, assuming an incoming plane wave.
However, nowadays available short and intense laser pulses interacting with matter create plasmas on a sub-laser period time scale \cite{gibbon,mulserbauer}. 
These plasmas, in turn, modify the further propagation of the laser pulse. We call this the extreme nonlinear optics regime, and in 
the case of (initially) spherical targets nonlinear Mie optics.

As the laser field propagation is determined by the electron density distribution 
and the plasma is generated by ionization, the charge state and density distributions is expected to be sensitive  to the ionization {\em dynamics}.  In fact, even the strongest present-day lasers cannot directly fully ionize heavier 
elements  so that the assumption of a preformed, throughout the target homogeneous plasma with a given dielectric constant may be inadequate. Furthermore, the skin-effect may prevent the laser 
from penetrating into targets that turn overdense in the course of ionization so that, in general, a richly structured space and 
time-dependent charge distribution develops \cite{enrique}. Such interactions of laser pulses with rapidly self-generated  plasmas 
have found already applications, e.g., as ``plasma mirrors,'' which are routinely used to increase the pulse contrast for intense 
laser-matter experiments \cite{PM-Kapteyn,PM-NatPhys}.

One expects that the part of the laser pulse that is  scattered off an overdense target will be mainly determined by the ratio of laser to plasma frequency at the surface whereas possible inhomogeneities inside the target do not play a role.  In fact, standard Mie scattering theory assuming a homogeneous, overdense plasma sphere was used to characterize rare gas clusters in recent experiments on a shot-to-shot basis \cite{bostedt,rupp}.
However, {\em inside} such a sphere, standard Mie theory would predict electric fields only within a narrow skin layer while,  in this Letter, we will show that   a highly inhomogeneous and temporally changing  charge density distribution may be created in the droplet interior. In order to probe such inhomogeneous structures inside the target laser frequencies greater than the plasma frequency that corresponds to the maximum plasma density in the target should be used.  Indeed, Thomson scattering of present-day short-wavelength free-electron laser radiation (from, e.g., DESY in Hamburg, LCLS in Stanford, or SACLA in Japan) is employed to probe overdense plasmas \cite{glenzer,sperlingI,sperlingII}.

 Molecular dynamics is a powerful tool that is widely used to describe the ionization dynamics 
in small laser-driven clusters \cite{doeppner,fennel_rev,mikaberidze}. 
However, for wavelength-sized targets such as droplets the influence of the target on the propagation of the incident electromagnetic wave needs to be taken into account self-consistently.
This requires the solution of Maxwell's equations together with the equations of motion for the charged particles. 
In the case of weakly coupled plasmas the problem can be reduced to the solution of the Vlasov-Maxwell system of equations, which 
is efficiently achieved using particle-in-cell 
(PIC) codes \cite{gibbon}.

Numerically, we study the non-linear Mie domain by means of a 3D relativistic PIC simulations with ionization included. 
The code UMKA originated from the study in \cite{code}. We show that in a certain laser intensity regime the droplet target is neither fully ionized nor are charges only created at the  
droplet surface. Instead, fields penetrate under a characteristic angle into the 
droplet, ionizing atoms in the polarization plane and triggering plasma waves which collide in a focal spot.  
We present results for the fractional ionization degree 
at various laser intensities, wavelengths, and densities, that turn out to follow an universal scaling law.

\begin{figure*}
\centerline{
\includegraphics[width=0.8\textwidth]{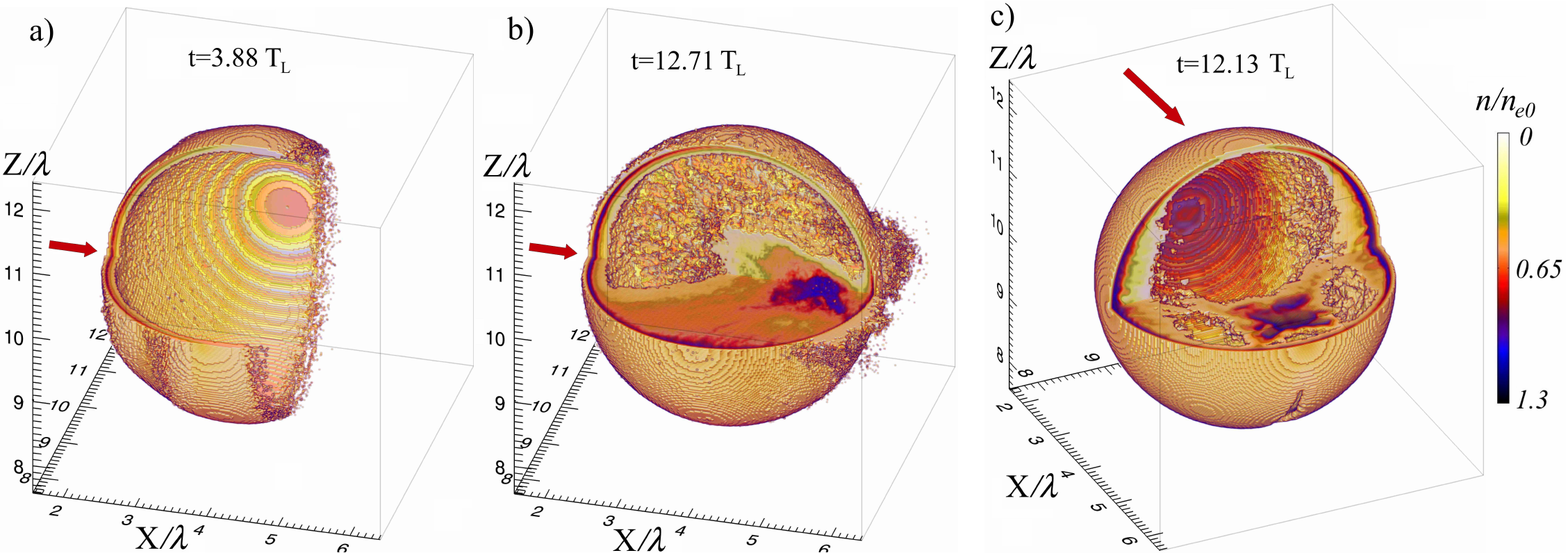}
}
\caption{(color online). Electron density 
in the beginning (a) and at the end (b) of the interaction with the laser pulse. He$^{2+}$ density (c)
at the end of the interaction. 
For better visualization of the droplet interior a quarter of it was cut-out. Laser and droplet parameters are given in the text. 
The laser propagation direction is indicated by an arrow in each panel.
\label{fig:He_3D}}
\end{figure*}

{\em Simulations ---} The ionization of an ion with charge state ${\cal{Z}}-1$  and ionization potential $I$ due to 
the electric field ${E}$ is implemented using 
the tunneling ionization rate formula \cite{popov04}
\begin{equation} w({E})=\left(\frac{2{E}_{ch}}{|{E}|}\right)^{2n^*}\frac{k^2\hbar}{m}\frac{|{E}|}{{E}_{ch}}
\exp\left(-\frac{2{E}_{ch}}{3 |{E}|}\right)\label{ionirate}\end{equation}
with
$k=\frac{\sqrt{2 m I}}{\hbar},\, {E}_{ch}= \frac{\hbar^2 k^3}{m e}, \,
n^*={\cal{Z}}\sqrt{\frac{I_H}{I}}. $
Here, $m$ is the electron mass and $I_H$ is the ionization potential of atomic hydrogen. 
When an ionization event takes place a free electron  at rest is created at the position of the ion. 
The energy needed for ionization is taken out of the field via an ``ionization current''  
$\mathbf{j}_\mathrm{ion}$ parallel to the electric field at the ion location. The value of $\mathbf{j}_\mathrm{ion}$
is such that $\mathbf{j}_\mathrm{ion}\cdot\mathbf{E}$ is the work spent on ionization per time step \cite{currentI,currentII}. 
Energy conservation is accounted for during the whole process; if the remaining field energy in a cell is insufficient 
for further ionization, this cell is not considered anymore during the current time step \cite{ruhl}.

We start by presenting typical results from PIC simulations of the interaction of an intense, plane-wave laser pulse with an initially 
neutral He droplet \cite{supplmat}. A spatial resolution of $\Delta x=\Delta y=\Delta z = \lambda/100,$ 125 macro-ions and 250 macro-electrons per cell were used.
Absorbing boundary conditions for the fields and particles were employed in the propagation direction, periodic ones for the other directions.
The size of the simulation box was always chosen big enough to rule out any boundary effects on the observables of interest due to reflections or particles leaving the box.
A linearly (in $y$-direction) polarized 10-cycle sin$^2$-laser pulse of carrier frequency $\omega_0$ enters the numerical box through the
boundary $x=0$ and propagates into the region $x > 0.$ The dimensionless vector potential 
amplitude $a=|e\hat{A}/m c|=|e\hat{E}/m\omega_0 c|$ was $0.5$, corresponding to a laser intensity 
$I\simeq 5.2\times 10^{17} \, $W/cm$^2$, the wavelength $\lambda=2\pi c/\omega_0$ was $800$\,nm (i.e. for a laser period $T_L=2.66$\,fs). 
The density of the $2R=4\lambda = 3.2 \mu$m diameter He-droplet was $\rho=0.14\, $g/cm$^{-3}.$ If the droplet was completely 
preionized such a density would correspond to an electron density $n_{e0}=24\, n_{cr},$ where $n_{cr}=1.8\times 10^{21}$cm$^{-3}$ is the critical density 
for $800$-nm wavelength light. The droplet center was located at $x=4\lambda$, $y=z=10\lambda$.
In the simulations presented in this Letter impact ionization was ``switched-off''. Test runs showed that for the intensities 
considered  the effect of collisional ionization during the laser pulse is much smaller than that of field ionization. 
Moreover, self-consistency requires that collisional absorption is taken into account along with the collisional ionization, 
as in the PIC codes described in Refs.~\cite{Sentoku_ion,OSIRIS,CALDER}.
A recently introduced microscopic PIC code \cite{MICPIC,peltz} bridges the gap between PIC and molecular dynamics and is 
also capable of incorporating collisional ionization and collisional absorption, albeit so far only for smaller targets.

\begin{figure}
\includegraphics[width=1.0\columnwidth]{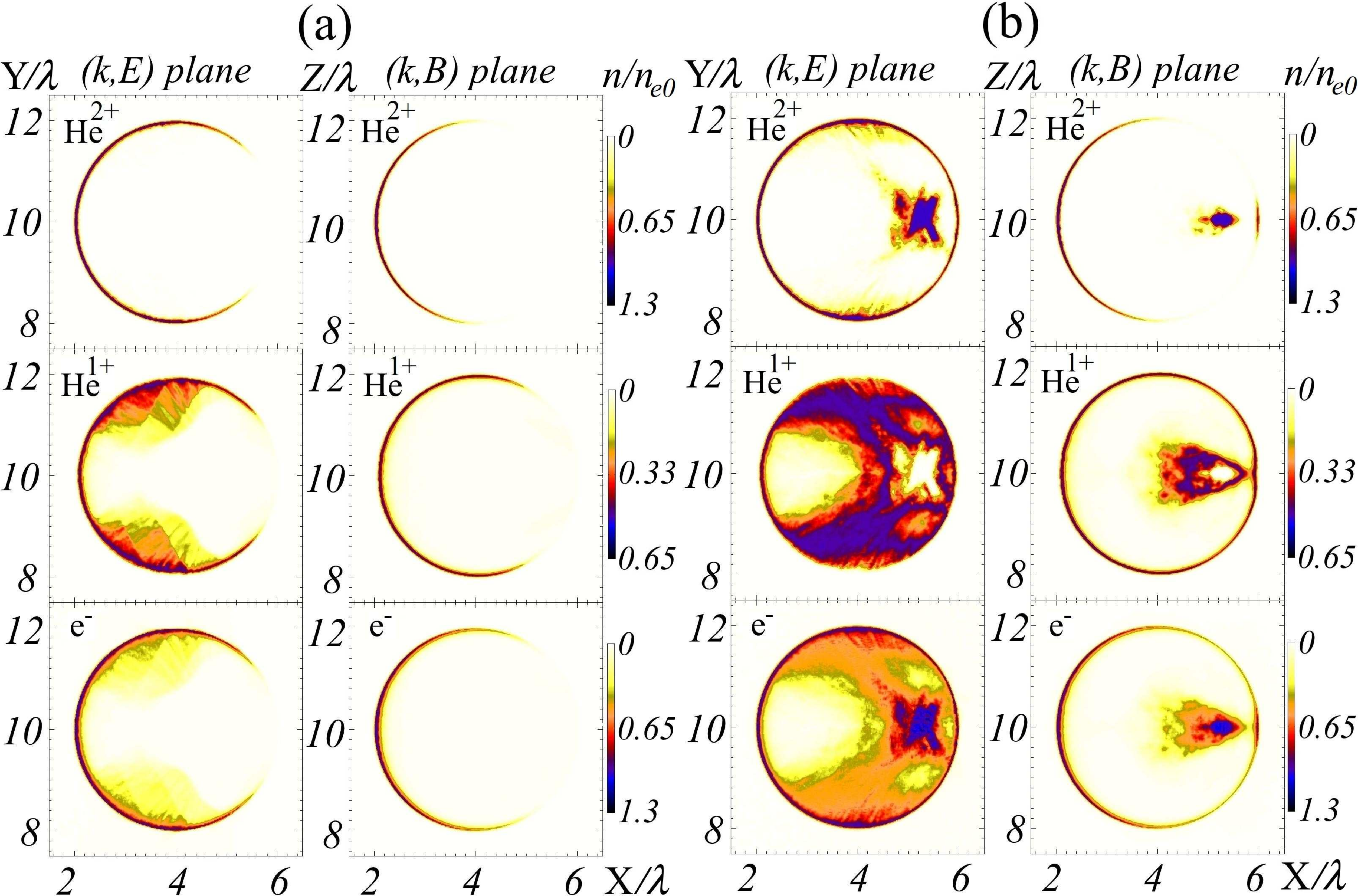}
\caption{(color online). Electron (bottom), He$^{1+}$ (middle) and He$^{2+}$ (top) density
in two perpendicular planes ($\mathbf{\hat{k}},\mathbf{\hat{E}}$) (left) and $(\mathbf{\hat{k}},\mathbf{\hat{B}})$ (right) at $t=7T_L$ (a) 
and $t=12T_L$ (b). 
Laser and droplet parameters are given in the text. 
 \label{fig:He_tunnel1}}
\end{figure}

{\em Results ---} Figure \ref{fig:He_3D} shows snapshots of the volume distribution of electron and He$^{2+}$ densities.  In the beginning the droplet is non-ionized and thus transparent for the leading part of the laser pulse.
Later, as the field strength of the laser pulse increases in magnitude, ionization becomes more efficient, and an overdense plasma 
is generated rapidly on the droplet surface as the pulse propagates over it, leading finally to almost full ionization  of a thin surface layer.  Moreover, we observe that a highly inhomogeneous density distribution inside the droplet is formed, concentrated 
mostly in the polarization plane. In particular, there seems to be a focal spot (blue area in the polarization plane in Fig.~\ref{fig:He_3D}b and \ref{fig:He_3D}c).
The fractional ionization degree $I_r=\frac{3}{4\pi R^3 n_{e0}}\int n_e(\mathbf{r})\,  d^3 r$ of the droplet at the end of the interaction is $\simeq 35\%$.

\begin{figure}
\includegraphics[width=1.0\columnwidth]{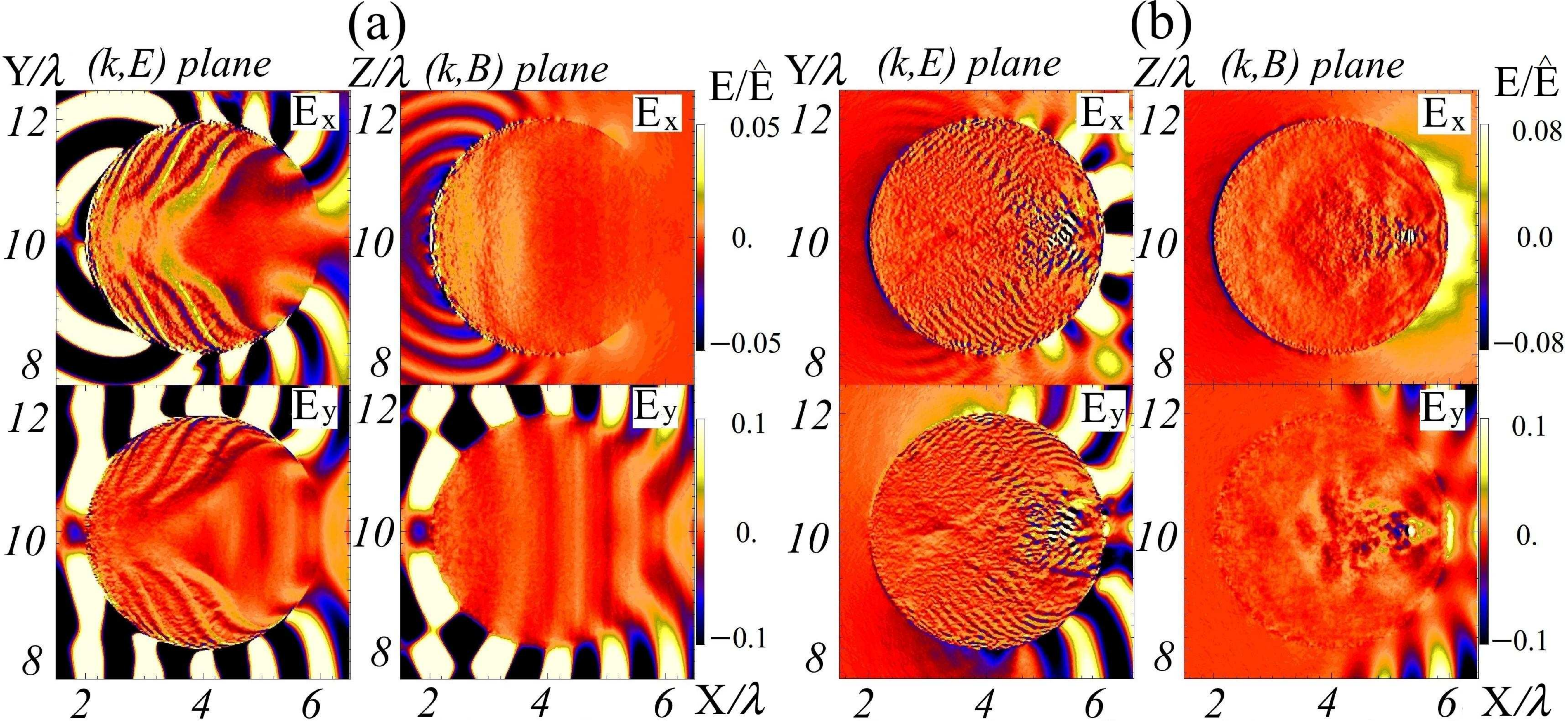}
\caption{(color online). Electric fields ($E_x$ top, $E_y$ bottom) in two perpendicular planes ($\mathbf{\hat{k}},\mathbf{\hat{E}}$) (left) 
and $(\mathbf{\hat{k}},\mathbf{\hat{B}})$ (right) at $t=7T_L$ (a) and $t=12T_L$ (b). 
\label{fig:He_tunnel2}} 
\end{figure}

He$^{1+},$ He$^{2+}$, and electron densities in the two perpendicular planes 
($\mathbf{\hat{k}},\mathbf{\hat{E}}$) and ($\mathbf{\hat{k}},\mathbf{\hat{B}}$) at times $t=7T_L$ and $t=12 T_L$ are plotted in Fig.~\ref{fig:He_tunnel1}. Charge density builds up inside the droplet, starting from a certain region on the droplet surface, most clearly seen in the He$^{1+}$-plot in Fig.~\ref{fig:He_tunnel1}a. At the later time in  Fig.~\ref{fig:He_tunnel1}b the charge density fronts merged already, creating the focal spot of  He$^{2+}$ density. Comparing the charge densities in the two planes shows that the ionization dynamics mainly takes place in the polarization plane  ($\mathbf{\hat{k}},\mathbf{\hat{E}}$). The corresponding distributions of the electric field components (longitudinal $E_x$ and in polarization direction of the incident laser electric field $E_y$)  are presented in Fig.~\ref{fig:He_tunnel2}. 
It is seen that an oscillating electric field penetrates into the droplet where in  Fig.~\ref{fig:He_tunnel1} the charge density is created. This is another interesting example for an electric field propagating in a plasma that is created by it in the first place \cite{Debayle}. More precisely, it turns out that the Mie-enhanced field  at the surface (discussed in the subsequent paragraph) first results in a deeper penetration and thus more efficient ionization. In addition, the electric field at the surface oscillates and thereby triggers plasma waves which propagate inwards up to the region where plasma has not yet been created. The electric field of the plasma wave then ionizes further, which results in an ionization front propagating inwards.

\begin{figure}[ht]
\centerline{
\includegraphics[width=1.0\columnwidth]{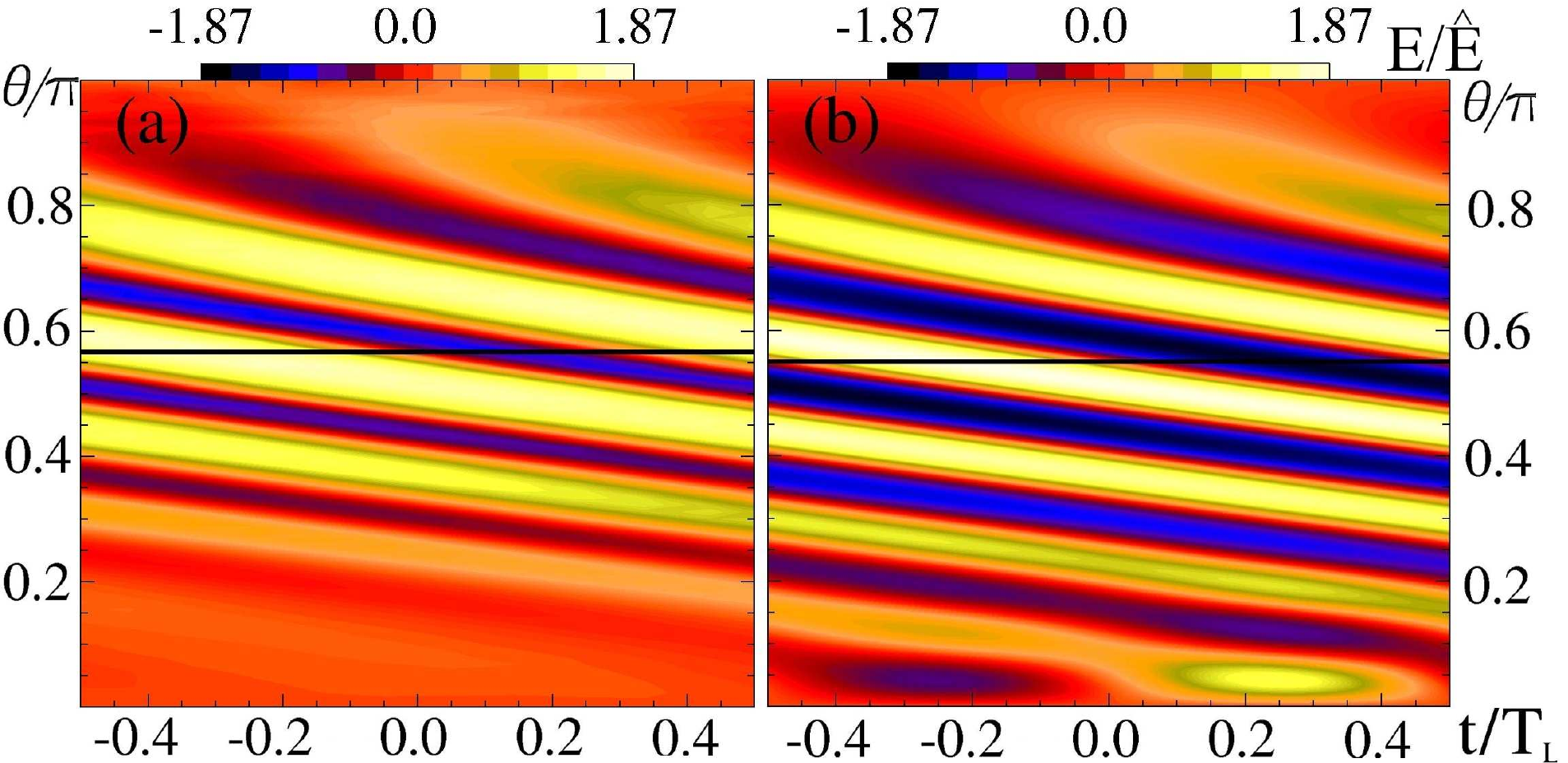}
}
\caption{(color online). Radial electric field along the surface in the polarization plane  vs angle $\theta$ and time,  obtained from the PIC simulation (a) 
and as predicted by Mie theory (b).  
The horizontal black lines indicate the angle at which the electric field at the droplet surface is highest.
\label{fig:Mie-Pic}} 
\end{figure}

{\em Mie field enhancement ---} We attribute the fact that the field and ionization front dynamics originate from a surface region under a certain angle $\theta\gtrsim \pi/2$ (with respect to $\mathbf{\hat{k}}$) to  a local, 
time-dependent field enhancement on the droplet surface.  In order to corroborate this statement, we show in Fig.~\ref{fig:Mie-Pic} the radial electric field along the droplet surface in the polarization plane vs time and $\theta$ as  obtained from the PIC simulation (a) and according to Mie theory \cite{Mie} (b). 
Standard Mie theory is formulated for plane incident waves.  As Mie theory is linear we synthesized our pulse via spectral decomposition and added the fields coherently.  In the Mie simulation the droplet is assumed to be homogeneous and conducting, with a dielectric constant $\epsilon=1-n_{e0}/n_{cr}$. Under such conditions Mie theory predicts in the strongly overdense regime (where  the skin  depth is $\delta_e\simeq c/\omega_p \ll R$ with $\omega_p=\sqrt{e^2 n_{e0}/m_e \epsilon_0}$ the electron plasma frequency) that the electric field on the droplet surface is perpendicular to it. 
In Fig.~\ref{fig:Mie-Pic} the time axis has been shifted such that  $t=0$ corresponds to the moment when the maximum of the incident laser pulse arrived at the droplet center. 
 Both PIC and Mie result predict maxima  of the electric field on the 
droplet surface for angles $\theta/\pi \in [0.4,0.7]$. The slight disagreement in the field distributions in forward direction (small $\theta$) is due to the fact that in the Mie calculation the droplet is assumed conducting (i.e., completely ionized) from the very beginning whereas in  the PIC simulation there is not yet plasma  at the rear side of the droplet  (see Fig.~\ref{fig:He_tunnel1}). The field enhancement predicted by Mie theory is in excellent agreement with the PIC results ($\simeq 1.9$ times the incident field).

\begin{figure}[ht]
\includegraphics[width=1.0\columnwidth]{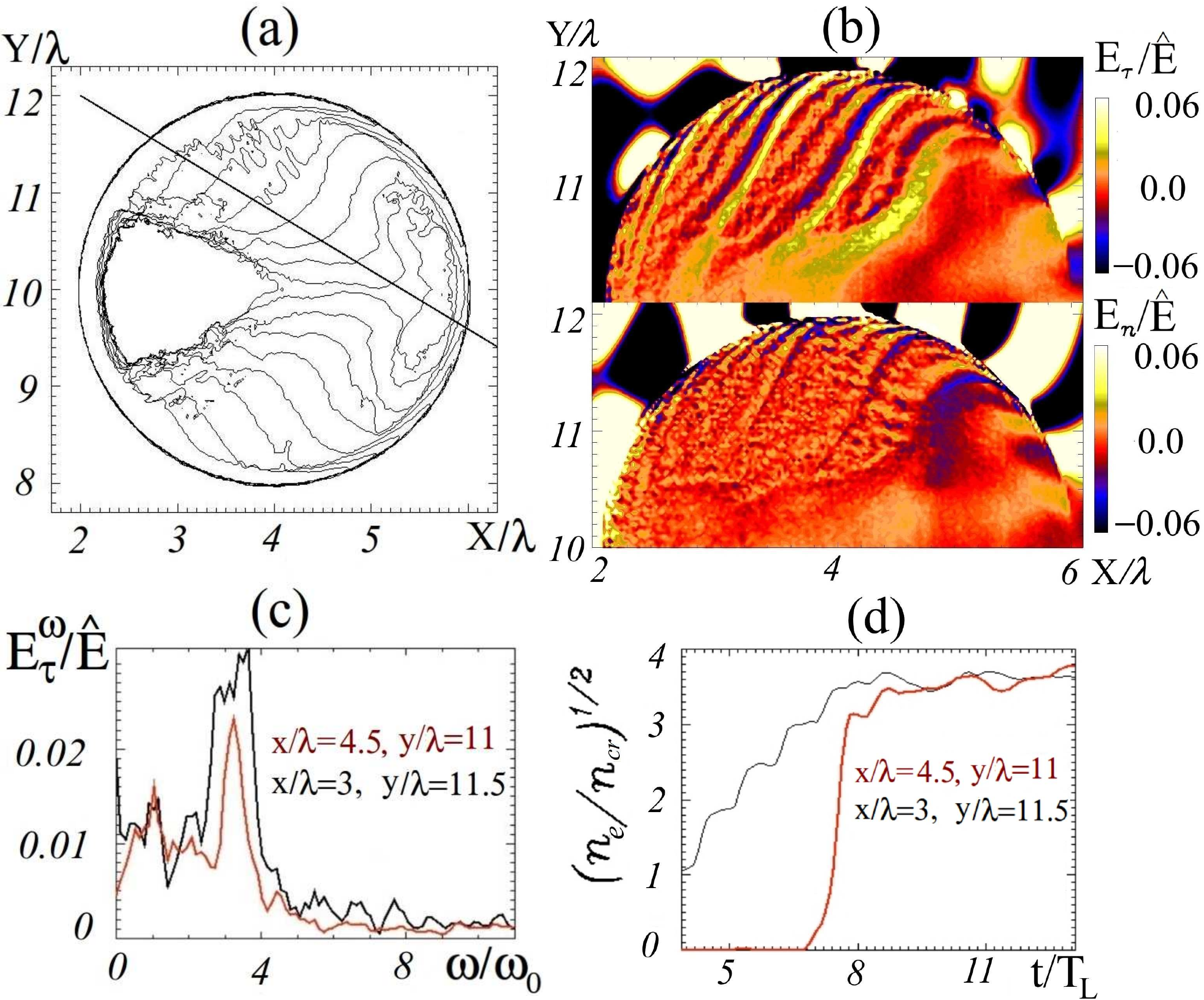}
\caption{(color online). (a) The $n=2.2\,n_{cr}$ level of the electron density in the droplet central plane $z=10$  for successive times ($t=5, 5.4, 6.1, 6.8, 7.4, 8.1, 8.8, 9.4, 10.1\, T_L$). 
The straight line indicates the projection direction $\boldsymbol\chi$. 
(b) The distributions of the electric field component  $E_\tau$ parallel 
and $E_n$ perpendicular to $\boldsymbol\chi$  at $t=8\,T_L$. 
(c) Frequency spectra of the electric field $E_{\tau}$ at the points $\mathbf{r}_1=(3,11.5,10)$ and $\mathbf{r}_2=(4.5,11,10)$ inside the droplet.
(d) $\sqrt{n_e/n_{cr}}$ vs time at the points  $\mathbf{r}_1$ and $\mathbf{r}_2$.  
\label{fig:Eper-Epar}}
\end{figure}

{\em Focused plasma waves ---} The propagation direction $\boldsymbol\chi$ of the field structures inside the droplet seen in  Fig.~\ref{fig:He_tunnel2}a is tilted with respect to $\mathbf{\hat{k}}$, leading to the observed focusing effect.  In order to interpret correctly these structures,  we project the field components inside the upper half of the droplet  onto $\boldsymbol\chi$,
$E_\tau=E_x\cos\varphi+E_y\sin\varphi,
\, E_n=-E_x\sin\varphi+E_y\cos\varphi,$ with $\varphi$ the angle between  $\boldsymbol\chi$ and $\mathbf{\hat{k}}$ (see Fig.\ref{fig:Eper-Epar}a).
The resulting field distributions for $E_\tau$ and $E_n$  are shown in Fig.~\ref{fig:Eper-Epar}b. As the values of $E_\tau$ are several times bigger than the values of $E_n$, we identify the field structures as a longitudinal plasma wave. The necessary matching of the plasma wave to the electromagnetic field at the droplet surface results in the tilt of  $\boldsymbol\chi$  with respect to $\mathbf{\hat{k}}$ because  the phase velocity of the plasma wave is smaller than $c$.   The frequency spectra of the electric field at two points $x=3\,\lambda$, $y=11.5\,\lambda$ and $x=4.5\,\lambda$, $y=11\,\lambda$   inside the droplet are shown in the Fig.~\ref{fig:Eper-Epar}c. They peak at the frequency approximately equal 
to the local plasma frequency, which may be estimated from the plot of the local density vs time in Fig.~\ref{fig:Eper-Epar}d.
 
\begin{figure}[ht]
\centerline{
\includegraphics[width=0.9\columnwidth]{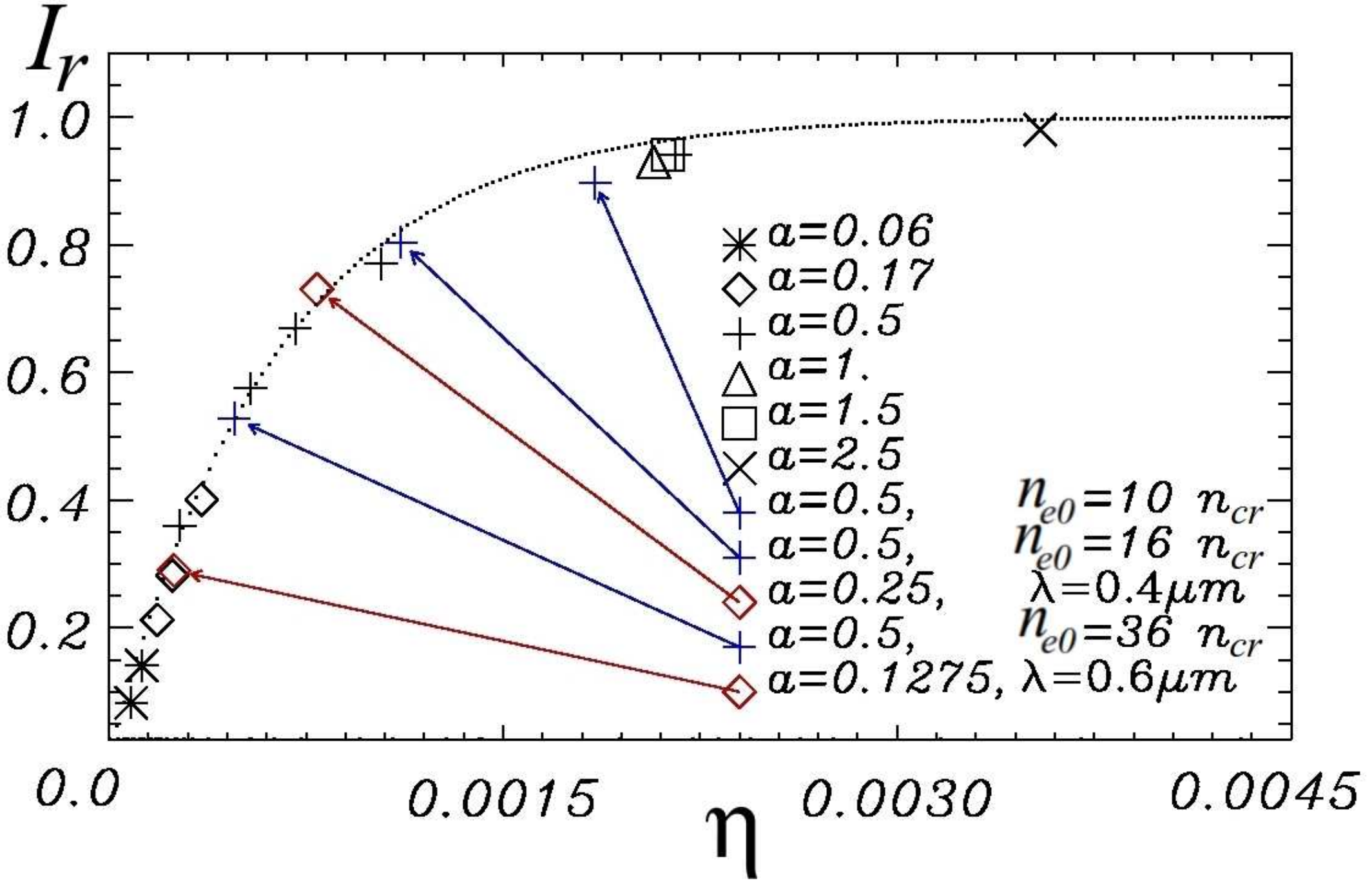}
}
\caption{(color online). Universal curve  $I_r=1-\exp(-\gamma \eta)$ (dotted) with $\eta=a/(R\lambda/\delta_e^2)=e\hat{E}/(2\pi mR (\omega_p^2-\omega_0^2))$ for the final fractional ionization degree of the He droplet after the interaction 
with a plane-wave laser pulse. The total laser pulse duration in all cases was $26$\,fs. The symbols for different $a$ are indicated in 
the plot. For some of them, runs with different density or laser wavelength have been performed, as indicated directly by arrows. 
The numerical values for $a$, $R/\lambda$, $n_{e0}/n_{cr}$, and $I_r$ are given in the Supplemental Material \cite{supplmat}. 
\label{fig:treshold}} 
\end{figure}

{\em Fractional ionization degree ---} Figure  \ref{fig:treshold} collects all our simulation results for  the final fractional ionization degree $I_r$. 
Introducing the dimensionless parameter  $\eta=a/(R\lambda/\delta_e^2)$,  it turns out that  for all the various cluster 
sizes $R>\delta_e$, densities $\omega_p^2\gg \omega_0^2$, laser intensities and wavelengths simulated,  $I_r$ is well 
described by  $I_r \simeq 1-\exp\left(-\gamma \eta\right)$. In our case of He we find $\gamma= 1560.$ Note that the species-dependence 
only enters via the ionization potentials $I$ in the tunneling ionization rate formula (\ref{ionirate}).    
Inserting the expression for the collisionless skin depth $\delta_e=c/\sqrt{\omega_p^2-\omega_0^2}$ we 
obtain $\eta=e\hat{E}/(2\pi mR (\omega_p^2-\omega_0^2))\simeq e\hat{E}/(2\pi mR \omega_p^2)$, showing that there is only a 
weak dependence on the laser frequency. Indeed, for tunneling ionization the electric field amplitude matters, not the laser frequency. 
For small laser intensity and sufficiently big droplets, when only the thin skin layer on the droplet surface gets ionized, 
one expects $I_r=\frac{4\pi R^2\delta_e}{4\pi R^3/3}\sim R^{-1}$. In the opposite limit of very high laser intensity or small droplets complete ionization $I_r =1$ is expected. Both limiting cases are contained in our formula. The chosen exponential interpolation between those two limiting cases matches the simulation results for  the fractional ionization degree very well.

{\em Summary ---} A strong near-infrared or optical laser pulse interacting  with an initially neutral, wavelength-sized 
He-droplet may generate  a charge density distribution that neither is homogeneous throughout the droplet nor created  only within 
a thin skin layer at the surface. Instead, electric fields may penetrate into the droplet interior  for certain angles of incidence 
predicted by standard Mie theory. However, the time-dependent field and density distributions inside the target are not accessible to standard Mie theory but fall into the realm of extreme nonlinear optics.  
 The field penetration causes ionization inside the droplet, mainly confined to the polarization plane. The resulting inhomogeneous charge distribution may be probed via scattering of short-wavelength radiation and should be taken into account when studying  typical  laser-plasma interaction applications such as ion acceleration or x-ray radiation from recombination in ionized droplets. A particularly high abundance of He$^{2+}$ is observed where the ionization fronts and the trailing plasma waves collide.
The fractional ionization degrees for various droplet and laser parameters are found to be in good agreement with a self-similar 
exponential fit. At higher laser intensities a qualitatively similar ionization dynamics is expected for higher-$Z$ materials as well.

\begin{acknowledgments}
This work was supported by the DFG within the SFB 652. PIC simulations were performed using the computing 
resources granted by the John von Neumann-Institut f\"ur Computing (Research Center J\"ulich) under the project HRO01. 
We thank Prof.\ Thomas Fennel for providing the result of the Mie calculation.
\end{acknowledgments}


\begin{thebibliography}{10}

\bibitem{Mie} Max Born and Emil Wolf, {\em Principles of Optics}
(Cambridge University Press, Cambridge, England, 2003).

\bibitem{gibbon} P. Gibbon, {\em Short Pulse Laser Interactions with Matter, an Introduction} (Imperial College Press, London 2005).

\bibitem{mulserbauer} P. Mulser and D. Bauer,
{\em High-Power Laser-Matter Interaction}
(Springer, Berlin Heidelberg, 2010).

\bibitem{enrique} E. C. Jarque, F. Cornolti, and A. Macchi, J. Phys. B: At. Mol. Opt. Phys. 33, 1
 (2000).

\bibitem{PM-Kapteyn} H. C. Kapteyn, M. M. Murnane, A. Szoke, and R. W. Falcone, Optics Lett. 16,  7 (1991).

\bibitem{PM-NatPhys} C. Thaury, F. Qu\'er\'e, J.-P. Geindre, A. Levy, T. Ceccotti, P. Monot, M. Bougeard, F. R\'eau, P. d'Oliveira, 
P. Audebert, R. Marjoribanks and Ph. Martin, Nature Physics 3, 424 (2007).

\bibitem{bostedt} C. Bostedt, E. Eremina, D. Rupp, M. Adolph, H. Thomas, M. Hoener, A.R.B. de Castro, J. Tiggesb\"aumker, K.-H. Meiwes-Broer, T. Laarmann, H. Wabnitz, E. Pl\"onjes, R. Treusch, J.R. Schneider, and T. M\"oller,
Phys. Rev. Lett. 108, 093401 (2012).

\bibitem{rupp} D. Rupp, M. Adolph, T. Gorkhover, S. Schorb, D. Wolter, R. Hartmann, N. Kimmel, C. Reich, T. Feigl, A.R.B. de Castro, R. Treusch, L. Str\"uder, T. M\"oller, and C. Bostedt, New J. Phys. 14, 055016 (2012).


\bibitem{sperlingII} P. Sperling, T. Liseykina, D. Bauer, R. Redmer, to appear in  New\ J.\ Phys. (March 2013).

\bibitem{glenzer} S.H. Glenzer, R. Redmer, Rev. Mod. Phys. 81, 1625 (2009).

\bibitem{sperlingI}   P. Sperling, R. Thiele,
    B. Holst,
    C. Fortmann,
    S.H. Glenzer,
    S. Toleikis,
    Th. Tschentscher,
    R. Redmer,
High Energy Density Phys. 7, 145 (2011).

\bibitem{doeppner} T. D\"oppner, J.P. M\"uller, A. Przystawik, S. G\"ode, J. Tiggesb\"aumker, K.-H. Meiwes-Broer, 
C. Varin, L. Ramunno, T. Brabec, T. Fennel, Phys. Rev. Lett. 105, 053401 (2010).

\bibitem{fennel_rev} Th. Fennel, K.-H. Meiwes-Broer,  J. Tiggesb\"aumker, P.-G. Reinhard, P.M. Dinh, E. Suraud, Rev.Mod. Phys. 82, 1793 (2010).

\bibitem{mikaberidze} A. Mikaberidze, U. Saalmann, J.M. Rost, Phys. Rev. Lett. 102, 128102 (2009).

\bibitem{code}  V.A. Vshivkov, N.M. Naumova, F. Pegoraro, and S.V. Bulanov, Phys. Plasmas 5, 2727 (1998).

\bibitem{popov04} V.S.\ Popov, Phys.\ Usp.\ 47, 855 (2004).


\bibitem{currentI} S. C. Rae, K. Burnett, Phys. Rev. A 46, 2077 (1992).
\bibitem{currentII} P. Mulser, F. Cornolti, D. Bauer, Phys. Plasmas 5, 4466 (1998).

\bibitem{ruhl} A. J. Kemp, Y. Sentoku, T. Cowan, J. Fuchs, H. Ruhl, Phys. Plasmas 11, L69 (2004).

\bibitem{supplmat} The case of focused pulses is discussed in the Supplemental Material ...

\bibitem{Sentoku_ion}  Y. Sentoku and A. J. Kemp, J. Comput. Phys. 227, 6846 (2008).

\bibitem{OSIRIS} R. A. Fonseca, L. O. Silva, F. S. Tsung, V. K. Decyk, W. Lu, C. Ren, W. B. Mori, S. Deng, 
S. Lee, T. Katsouleas, J. C. Adam, Lecture Notes in Computer Science V. 2331, pp 342-351   (2002).

\bibitem{CALDER} E. Lefebvre, N. Cochet, S. Fritzler, V. Malka, M.-M. Aleonard, J.-F. Chemin, S. Darbon, L. Disdier,
J. Faure, A. Fedotoff, O. Landoas, G. Malka, V. Meot, P. Morel, M. Rabec Le Gloahec, A. Rouyer, Ch. Rubbelynck,
V. Tikhonchuk, R. Wrobel, P. Audebert and C. Rousseaux, Nucl. Fusion 43,629  (2003).

\bibitem{MICPIC} Ch. Varin, Ch. Peltz, Th. Brabec, Th. Fennel, Phys. Rev. Lett. 108, 175007 (2012).

\bibitem{peltz} Christian Peltz, Charles Varin, Thomas Brabec and Thomas Fennel, New J. Phys. 14, 065011 (2012).

\bibitem{Debayle} A. Debayle and V. T. Tikhonchuk, Phys. Plasmas 14, 073104 (2007).





\end{thebibliography}
\end{document}